\documentclass[preprint]{aastex}
\usepackage{graphicx}
\makeatletter
\shorttitle{Helioseismic response to X2.2 solar flare of February 15, 2011}
\shortauthors{Kosovichev}
\bibpunct[, ]{(}{)}{; }{a}{,}{,}

\makeatother

\begin{document}

\title{Helioseismic response to X2.2 solar flare of February 15, 2011}

\author{A.G. Kosovichev}

\affil{ W.~W.~Hansen Experimental Physics Laboratory, Stanford University,
Stanford, CA 94305, USA }
\begin{abstract}
The X2.2-class solar flare of February 15, 2011, produced a powerful  
`sunquake' event, representing a helioseismic response to the flare 
impact in the solar photosphere, which was observed with the HMI 
instrument on the Solar Dynamics Observatory (SDO). The impulsively 
excited acoustic waves formed a compact wavepacket traveling through 
the solar interior and appearing on the surface as expanding wave 
ripples.  The initial flare impacts were observed in the form of 
compact and rapid variations of the Doppler velocity, line-of-sight 
magnetic field and continuum intensity. These variations formed a 
typical two-ribbon flare structure, and are believed to be 
associated with thermal and hydrodynamic effects of high-energy 
particles heating the lower atmosphere.  The analysis of the SDO/HMI 
and X-ray  data from the Ramaty High Energy Solar Spectroscopic 
Imager (RHESSI) shows that the helioseismic waves were initiated by 
the photospheric impact in the early impulsive phase, observed prior 
to the hard X-ray ($50-100$ keV) impulse, and were probably 
associated with atmospheric heating by relatively low-energy 
electrons ($\sim 6-50$ keV) and heat flux transport. The impact 
caused a short motion in the sunspot penumbra prior to the 
appearance of the helioseismic wave. It is found that the 
helioseismic wave front traveling through a sunspot had a lower 
amplitude and was significantly delayed relative to the front 
traveling outside the spot. These observations open new perspectives 
for studying the flare photospheric impacts and for using the 
flare-excited waves for sunspot seismology.
\end{abstract}


\section{Introduction}

`Sunquakes', the helioseismic response to solar flares, are
caused by strong localized hydrodynamic impacts in the photosphere
during the flare impulsive phase. The helioseismic waves are
observed directly as expanding circular-shaped ripples on the
solar surface, which can be detected in Dopplergram movies and as
characteristic ridges in time-distance diagrams,
\citep{Kosovichev1998, Kosovichev2006a, Kosovichev2006b}, or indirectly by
calculating the distribution of an integrated acoustic emission \citep{Donea1999,
Donea2005}.

Solar flares are sources of high-temperature plasma, strong 
hydrodynamic motions and heat flux in the solar atmosphere. Perhaps, 
in all flares such perturbations generate acoustic waves tra\-veling 
through the interior. However, only in some flares is the impact 
sufficiently localized and strong to produce the seismic waves with 
the amplitude above the convection noise level. The sunquake events 
with expanding ripples are relatively rare, and { have been} 
observed only  in some high-M and  X-class flares. The last previous 
observation of the seismic waves was reported for X1.2 flare of 
January 15, 2005.

It has been found in the initial July 9, 1996, flare observations
\citep{Kosovichev1998} that the hydrodynamic impact in the 
photosphere (`sunquake source') followed the hard X-ray flux 
impulse, and hence, the impact of high-energy electrons. { They} 
suggested that the mechanism of sunquakes can be explained by a 
hydrodynamic `thick-target' model of solar flares 
\citep{Kostiuk1975}. Several other mechanisms { of the helioseismic 
response}, including impact by high-energy protons and back-warming 
radiation heating \citep[e.g.][]{Donea2005,Zharkova2007}, and also 
due to magnetic field variations \citep{Hudson2008}. However, the 
mechanism, which converts a part of the flare energy and momentum 
into the helioseismic acoustic waves, is currently unknown. It is 
also unknown why only some flares generate { large-amplitude waves 
observed as ripples or enhanced acoustic emission}  \citep[see a 
review of ][for a recent discussion and references]{Hudson2011}.

Most of the previous observations of sunquakes were obtained with
the Michelson Doppler Imager instrument on SOHO. However, the 
full-disk observations with the full 2 arcsec/pixel resolution 
suitable for flare studies were obtained uninterruptedly only for 2 
months a year. Thus, many flares were not observed, and the 
statistics of sunquakes and their relation to the flare properties 
were not established. Except short eclipse periods in March and 
September, the Solar Dynamics Observatory launched in February 2010 
provides uninterrupted observations of the Sun. The Helioseismic and 
Magnetic  Imager (HMI) on SDO observes variations of intensity, 
magnetic field and plasma velocity (Dopplergrams) on the surface of 
Sun almost uninterruptedly with high spatial resolution (0.5 
arcsec/pixel) and high cadence (45 sec) \citep{Schou2010}. The flare 
of February 15, 2011, was the first X-class flare of the new solar 
cycle, 24, and the first  observed by HMI. This paper presents 
results of the initial analysis, which revealed the sunquake event. 
This event shows some curious properties, which make it different 
from the previously observed `sunquakes'.

\section{Results}

\begin{figure}
\begin{center}
\includegraphics[width=0.7\textwidth]{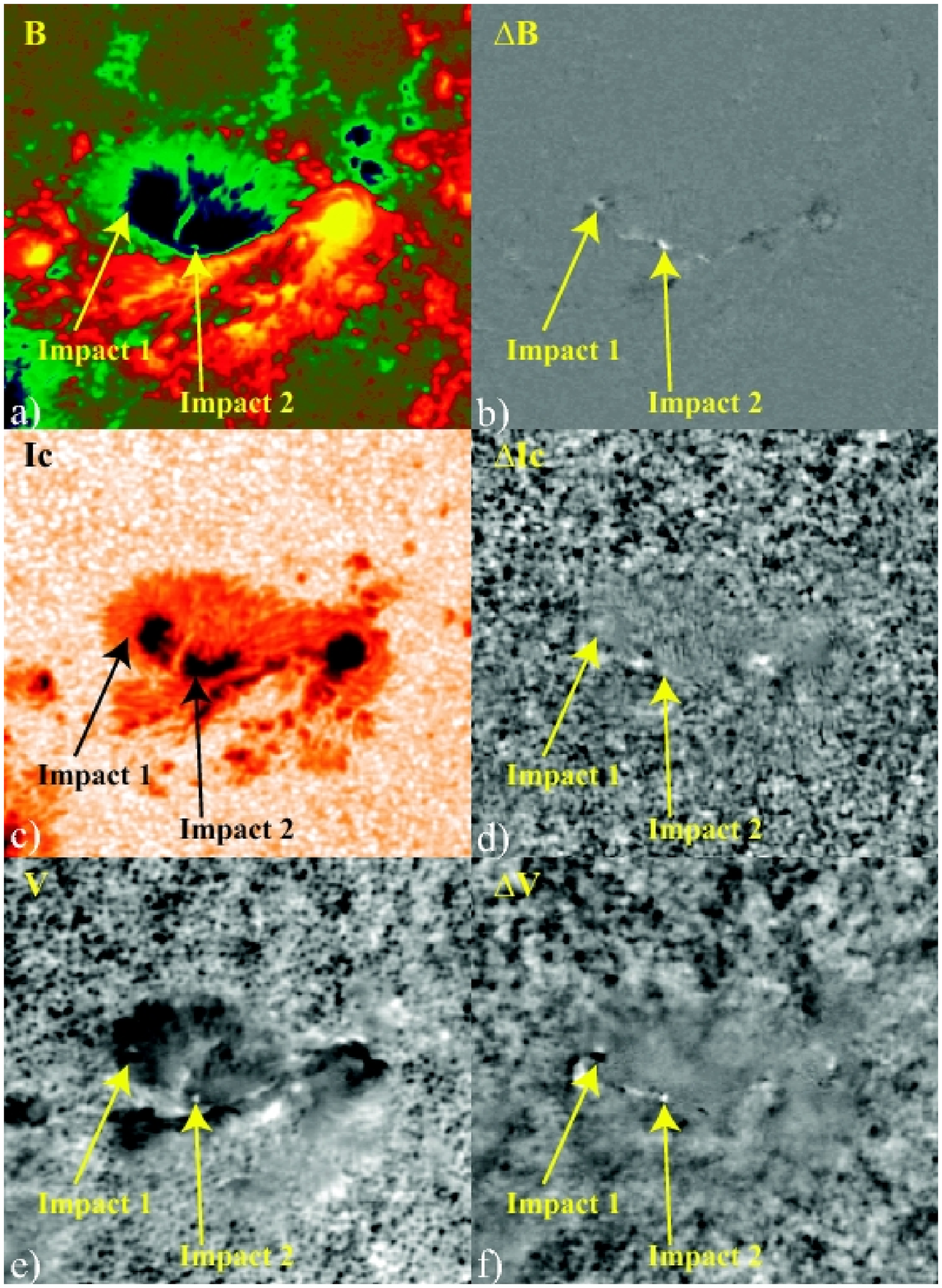}
\end{center}
\caption{a) Images of AR 1158 during the impulsive phase of the X2.2 flare,
 at 2011.02.15~ 01:54:39~TAI: a) line-of-sight magnetic field $B$, 
c) continuum intensity $I_c$, and e) Doppler velocity $V$.  
The right panels show the differences between these images and 
the corresponding images taken 45 sec earlier. The range of 
the magnetogram color map is $\pm 1$~kG; the range of the Dopplergram 
is $\pm 1$~km/s. Arrows show positions of two analyzed sources 
of transient flare variations located along the same flare ribbon. 
Traces of the second ribbon can be seen in panel b) to the right 
and below of `Impact 2'. A powerful sunquake originated from `Impact 1'. 
`Impact 2' is a place of a strong impulsive impact, but it did 
not generate a significant sunquake.}\label{fig1}
\end{figure}

The X2.2 flare of February 15, 2011, occurred  in the central 
sunspot of active region NOAA 1158, which had a $\delta$-type 
magnetic configuration (Fig.~\ref{fig1}). According to the GOES-15 
soft X-ray measurements, the flare started at 01:44~UT, reached 
maximum at 01:56~UT and ended at 02:06~UT. The flare signals are 
clearly detected in all HMI observables, and show that the flare had 
a typical two-ribbon structure with the ribbons located on both 
sides of the magnetic neutral line. This is  well seen in the 
magnetograms (Fig.~\ref{fig1}a). { We consider two places of strong 
localized photospheric impacts as `Impact 1' and `Impact 2' for 
further analysis. The sunquake originated from `Impact 1'.}

\begin{figure}
\begin{center}
\includegraphics[width=0.7\textwidth]{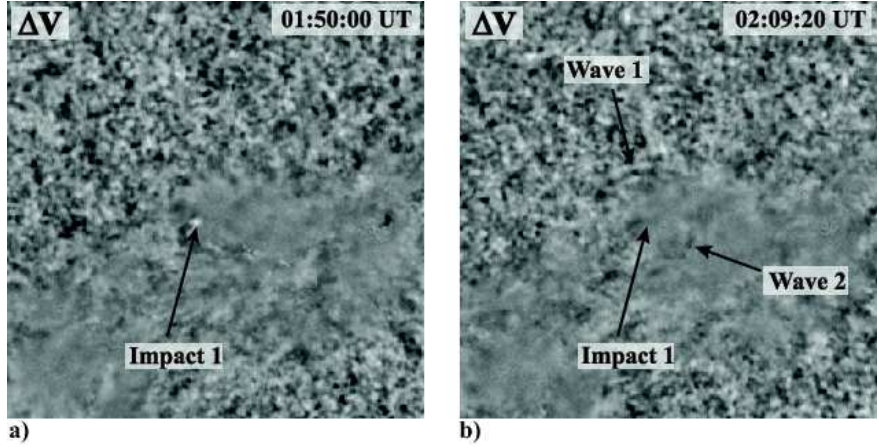}
\end{center}
% \begin{minipage}{\textwidth}
%\vspace*{-7mm}
\caption{Frequency-filtered Doppler velocity differences
, $\Delta V$,
 at the moments of: a) the flare impact at `Impact 1', calculated 
for 01:50:00 UT, and b) about 19 min later at 02:09:20 UT, showing 
the sunquake wavefront. These are two frames of the supplementary on-line movie. 
The movie is produced by interpolating the 45-sec cadence data into a new 
series with 20-sec cadence starting at 01:40~UT. This makes the movie slower 
and easier to watch. The original 45-sec cadence movie is available on-line 
in a RHESSI Science Nuggets article \citep{Kosovichev2011a}.}\label{fig2}
%\end{minipage}
\end{figure}

\begin{figure}
\begin{center}
\includegraphics[width=\textwidth]{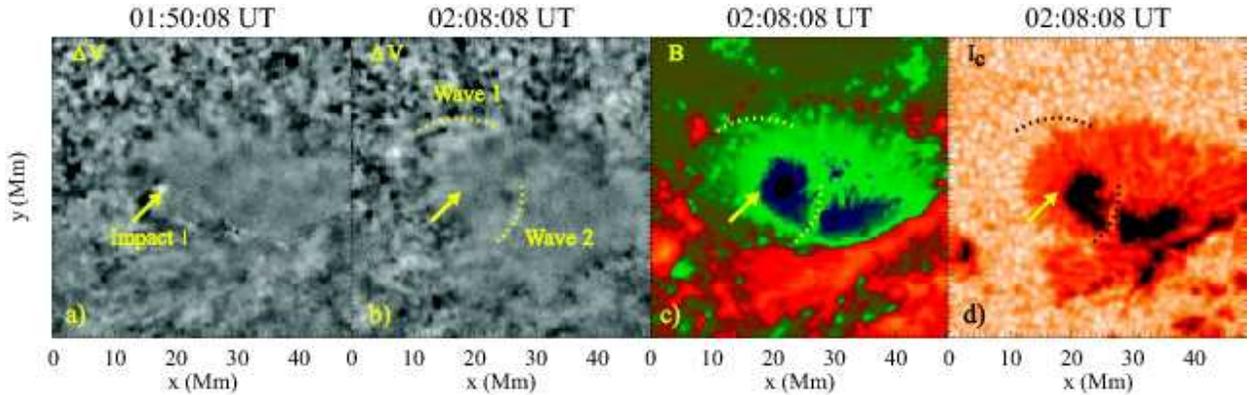}
\end{center}
%\begin{minipage}{\textwidth}
\caption{Illustration of positions of the flare `Impact 1' (panel a) 
and the helioseismic fronts (b) observed in the Doppler-shift data, 
$\Delta V$, in the corresponding magnetogram, $B$,  (c) and 
continuum intensity, $I_c$, (d) images. }\label{fig3}
%\end{minipage}
\end{figure}

The sunquake event was initially revealed  in the running difference 
movie of the raw Doppler velocity data \citep{Kosovichev2011a}. 
However, the wave structure is better seen after applying to the 
data a Gaussian frequency filter with a central frequency of 6~mHz 
and a characteristic width of 2~mHz. This filter enhances the 
high-frequency sunquake signal relative to the lower-frequency 
background solar granulation noise. In addition, the images are 
remapped onto the heliographic Carrington coordinates { using the 
Postel's azimuthal equidistant projection} and tracked with the 
differential rotation rate. Figure~\ref{fig2} shows two frames of 
the frequency-filtered Doppler-velocity movie { (available as 
supplementary online-only material)}. The sunquake wave appears 
about 20 minutes after the initial flare impact of the photosphere. 
The wave front has a circular shape, but it is not isotropic. The 
wavefront  traveling outside the magnetic region in the North-East 
direction (`Wave 1') has the highest amplitude, and is most clearly 
visible. In the opposite direction the wave travels through a 
sunspot (`Wave 2'), and its amplitude is suppressed. Also, the wave 
front traveling through the sunspot is visibly delayed relative to 
`Wave 1' traveling outside. Figure~\ref{fig3} shows positions of the 
two wave fronts at 02:08~UT in the corresponding magnetogram and 
white-light images.

\begin{figure}
\begin{center}
\includegraphics[width=0.8\textwidth]{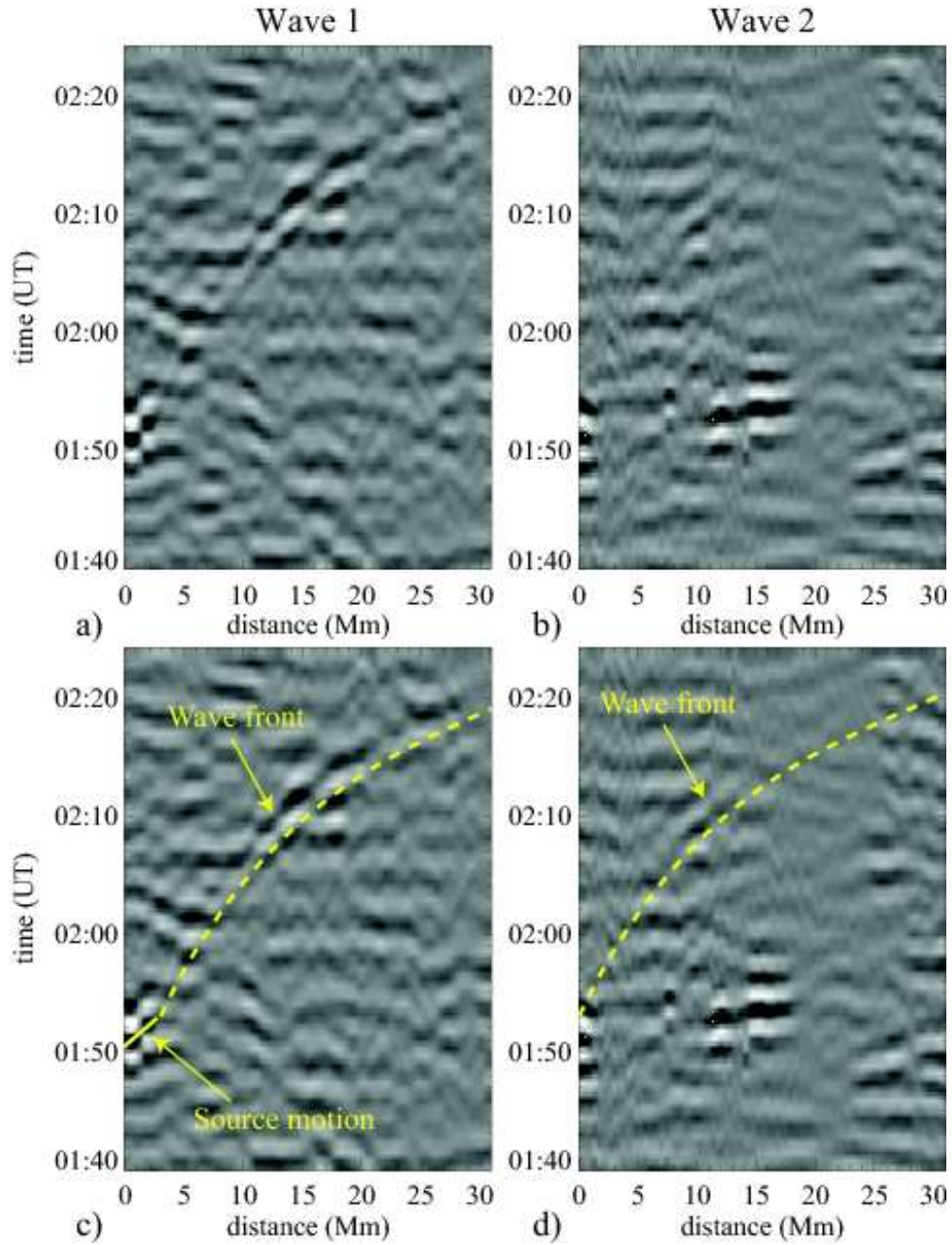}
\end{center}
\caption{Time-distance diagrams for: a) `Wave 1' and b) `Wave 2', 
both originating from `Impact 1'. In the duplicated diagrams c) and d), 
the dashed curves are the theoretical time-distance calculated for 
a standard solar model in the ray approximation. The locations of these 
curves are chosen to approximately match the leading wave fronts. 
The short solid line in panel c) indicates the source motion.}\label{fig4}
\end{figure}

Figure \ref{fig4} shows the time-distance diagrams  obtained by 
remapping the frequency-filtered Dopplergrams onto the polar 
coordinates with the center at `Impact 1', and averaging over the 
range of angles corresponding to the two parts of the wave fronts in 
Fig.~\ref{fig2}. In these diagrams, the helioseismic acoustic waves 
form characteristic ridges, the slope of which corresponds to the 
local group-speed of the wave packets traveling between two surface 
points through the interior. The speed increases with the distance 
because for larger distances the waves travel through the deeper 
interior where the sound-speed is higher 
\citep[e.g.][]{Kosovichev2011b}. For comparison, the theoretical 
travel times calculated in the ray approximation are shown by dashed 
curves. The starting points of these curves are chosen to 
approximately match the position of the ridges. Evidently, the ridge 
of `Wave 2' is much weaker and shorter than the ridge of `Wave 1'. 
In these diagrams, the wave source (`Impact 1') produces strong 
variations at zero distance at about 01:50~UT. Also, the `Wave 2' 
front is delayed by $\sim 100$ sec with respect to `Wave 1'.

The time-distance diagrams show two interesting features.  During 
approximately the first three minutes the wave source is moving in 
the direction of `Wave 1' (Fig.~\ref{fig4}) with a speed of about 
15-17~km/s, which is higher than the local sound speed but may 
correspond to the  magneto-acoustic speed of the sunspot penumbra in 
the vicinity of the source. { The source motions, which can be 
supersonic,} have been observed for other `sunquake' events 
\citep{Kosovichev2006b,Kosovichev2006c,Kosovichev2007}. Similar to 
this case, the source motions are at least partly responsible for 
the anisotropy of the wave amplitude. In the previous cases the 
source motion was associated with apparent motions of the point-like 
photospheric impacts in the flare ribbons. In this case, the source 
motion may be associated with MHD waves excited by the flare 
momentum impact in the almost horizontal field on the penumbra. 
However, this process requires a special separate investigation.

\begin{figure}
\begin{center}
\includegraphics[width=0.9\textwidth]{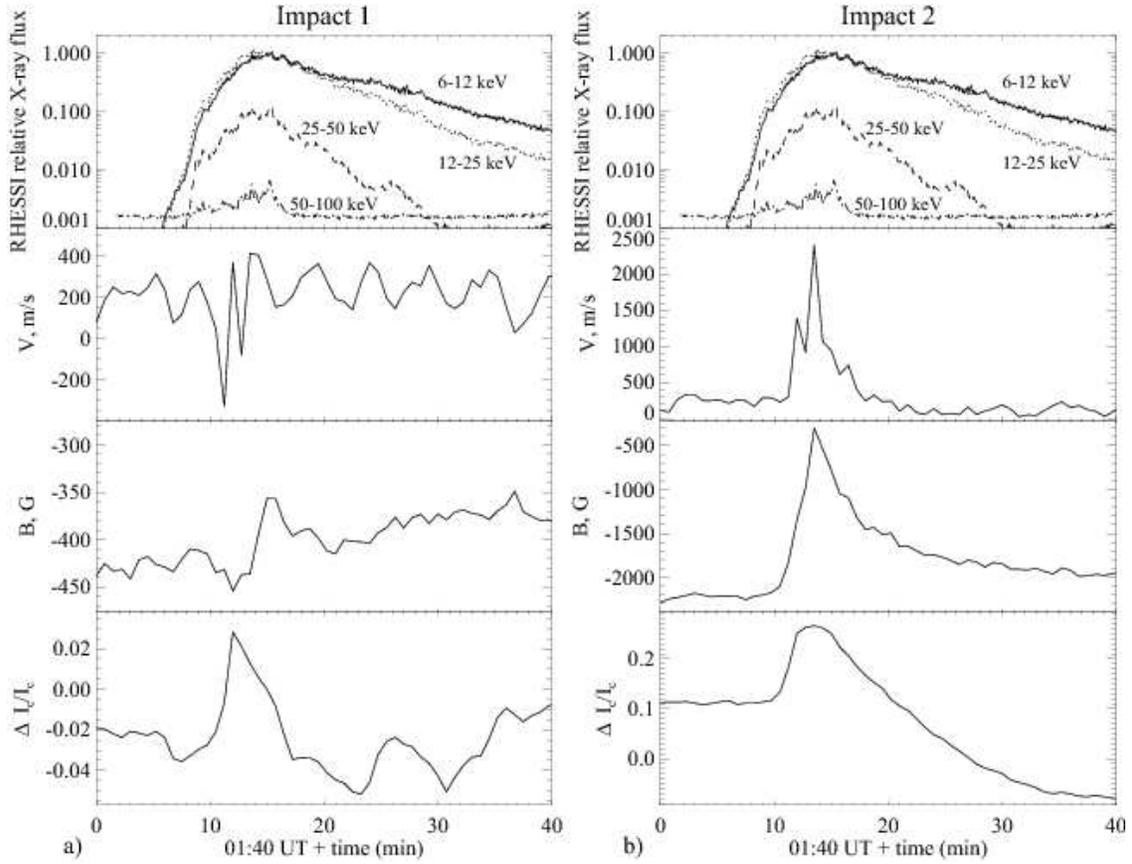}
\end{center}
\caption{Variations of the total X-ray fluxes from RHESSI 
(relative to the 6-12 keV flux), Doppler velocity, magnetic field 
and continuum intensity at: a) `Impact 1'; b) `Impact 2'. 
The X-ray fluxes are integrated for the whole flare region, 
and thus are identical in panels a) and b).}\label{fig5}
\end{figure}

The sunquake source is associated with one of the impacts located 
along the flare ribbons. The flare ribbons consist of individual 
patches representing impacts of flare impulses. In these data it is 
easy to find that the location of the sunquake source was in the 
penumbra area near the edge of the active region. This area is 
identified as `Impact 1' in Figures~\ref{fig1}--\ref{fig3}. It is 
characterized by strong and rapid variations of the Doppler velocity 
and magnetic field, and also by an impulsive increase of the 
continuum intensity (Fig.~\ref{fig5}a). There were strong 
photospheric impacts in several other locations. However, these 
impacts did not provide clearly visible seismic waves. The reason 
for this is not clear. For comparison, in Figure~\ref{fig5}b we show 
the  variations in one of the strongest compact impacts identified 
as `Impact 2'. This impact was located in a region of strong  
magnetic field in the sunspot outer umbra near the magnetic neutral 
line. During the impact, the HMI data show a strong increase of the 
Doppler velocity, indicating downflows, a sharp impulsive decrease 
of the magnetic field strength, which relaxed to a value lower the 
the pre-flare strength, and an increase in the continuum intensity 
brightness. All the variations in `Impact 2' are stronger than in 
`Impact 1', but this impact did not generate strong sunquake 
ripples. It seems that the main difference between these two places 
of the flare impact onto the photosphere of the Sun is that `Impact 
1' was located in a region of relatively weak ($\sim 400~$G) 
magnetic field contrary to `Impact 2', which was in strong field 
($\sim 2000~$G). In addition, `Impact 1' was more variable and 
moving. It started near the inner boundary of the penumbra (bright 
point at the `Impact 1' arrow in Fig.~\ref{fig3}a) and then moved 
into the penumbra, generating a localized motion in this part of the 
penumbra. The dynamic nature of the flare impact seems to be 
important for understanding the mechanism of sunquakes. The strong 
magnetic field in `Impact 2' probably restricted wave motions, and, 
perhaps, this may explain the absence of helioseismic response from 
this impact.

The origin of photospheric impacts during the flare impulsive phase 
is yet to be understood. In this case, it is particularly puzzling 
that the initial `Impact 1' occurred in the early impulsive phase, 
prior to the hard X-ray  impulse in the energy range of 50-100 keV, 
and just at the beginning of the X-ray 25-50 keV impulse 
(Fig.~\ref{fig5}a). The traditional `thick-target' mechanism of the 
energy transport in solar flares \citep[e.g.][]{Hudson2011} assumes 
that most of the energy is released in the form of high-energy 
electrons, which heat the solar chromosphere generating a localized 
high-pressure zone. This zone explodes, and causes `chromospheric 
evaporation' into the corona and the hydrodynamic impact in the 
photosphere, which leads to `sunquake'. However, in this case the 
photospheric impact apparently happened  before the main particle 
acceleration phase. This requires a new mechanism of the energy and 
momentum transport into the low atmosphere during the early 
`pre-heating' flare phase.   Certainly, further investigations of 
the sunquake events, their energetics and dynamics, will provide new 
insight in the mechanisms of the flare energy release and transport.

\section{Discussion}

The first observations of the sunquake event from SDO/HMI revealed 
very interesting properties of the flare impact onto the solar 
photosphere. The HMI data with the significantly higher resolution  
than the previous SOHO/MDI observations of sunquakes  provide a new 
insight into the dynamics of the flare impact and the sunquake 
source. The preliminary analysis indicates that seismic flare waves 
are generated by the impact in the region of a relatively weak 
magnetic field of the sunspot penumbra. A significantly stronger 
impact in a region of high magnetic field strength did not generate 
helioseismic waves of a comparable magnitude.

A characteristic feature of this sunquake is anisotropy of the wave 
front: the observed wave amplitude is much stronger in one direction 
than in the others. This was observed also in previous events. In 
particular, the seismic waves excited during the October 28, 2003, 
flare had the greatest amplitude in the direction of the expanding 
flare ribbons. The wave anisotropy was attributed to the moving 
source of the hydrodynamic impact, which is located in the flare 
ribbons \citep{Kosovichev2006b,Kosovichev2006c}. The motion of flare 
ribbons is often interpreted as a result of the magnetic 
reconnection processes in the corona. When the reconnection region 
moves up it involves higher magnetic loops, the footpoints of which 
are further apart. This may explain the expanding flare ribbons (as 
places of the photospheric flare impacts) and the association of 
sunquakes with the ribbon sources. In this event, the sunquake had a 
similar dynamical property: it started at an inner boundary of the 
sunspot penumbra and then quickly moved in the penumbra region. This 
was accompanied by a motion of this part of the penumbra. This is 
certainly an interesting phenomenon, which requires further 
investigation. Of course, there might be other reasons for the 
anisotropy of the wave front, such as inhomogeneities in 
temperature, magnetic field and plasma flows. However, the source 
motion seems to be quite important for generating sunquakes.  In 
addition, the wave front traveling through the sunspot umbra is 
significantly delayed relative to the wave front traveling outside 
the sunspot. This delay may be related to the source motion and also 
to a lower wave speed in the sunspot umbra. Theoretical MHD modeling 
of the dynamic impact source and the wave propagation in sunspot 
models is necessary for the understanding of this phenomenon.

The comparison of the SDO/HMI observations with the X-ray  
observations from RHESSI shows that the photospheric impact, which 
led to the excitation of the helioseismic waves, occurred at the 
beginning of the flare impulsive phase, before the hard X-ray 
impulse in the energy range of 50-100 keV and before main particle 
acceleration phase. { Perhaps,} the energy transport into the lower 
atmosphere may be provided by the saturated heat flux as recently 
suggested { for chromospheric evaporation} by \citet{Battaglia2009}. 
Theoretical models of the heat flux-saturated (or flux-limited) 
energy transport in the solar atmosphere were previously studied by 
several authors \citep[e.g.][]{Smith1986,Karpen1987}. 
\citet{Kosovichev1988} showed that this transport has wave 
properties with a sharp shock-like heat front. Alternatively, the 
difference in timing between the photospheric impact and the hard 
X-ray flux may be related to changes in anisotropy of high-energy 
electrons during the flare impulsive phase \citep[V. Petrosian, 
private communication; e.g.][]{Leach1985}. It will be important to 
further investigate the role of high-energy particles, thermal and 
MHD effects in the initial phase of solar flares.

%\bibliography{rhessi2011}


\end{document}